\documentclass[12pt]{article}
\usepackage{epsfig,amssymb}

\begin{document}
\pagenumbering{arabic}
\thispagestyle{empty}
\def\thefootnote{\fnsymbol{footnote}}
\setcounter{footnote}{1}

\begin{flushright}LC-TH-2002-005
\end{flushright}

\vspace{2cm}

\begin{center}
{\Large\bf Logarithmic Fingerprints of Virtual Supersymmetry
\footnote{Partially supported by EU contract HPRN-CT-2000-00149}}
 \vspace{2.5cm}  \\
%-----------------------------------------------------------------
{\large M. Beccaria$^{a,b}$, 
F.M. Renard$^c$ and C. Verzegnassi$^{d, e}$}\\
\vspace{0.7cm}
$^a$Dipartimento di Fisica, Universit\`a di
Lecce,
Via Arnesano, 73100 Lecce, Italy.\\
\vspace{0.2cm}
$^b$INFN, Sezione di Lecce\\
\vspace{0.2cm}
$^c$ Physique
Math\'{e}matique et Th\'{e}orique, UMR 5825\\
Universit\'{e} Montpellier
II,  F-34095 Montpellier Cedex 5.\hspace{2.2cm}\\
\vspace{0.2cm}
$^d$
Dipartimento di Fisica Teorica, Universit\`a di Trieste, 
Strada Costiera
 14, Miramare (Trieste) \\
\vspace{0.2cm}
$^e$ INFN, Sezione di Trieste\\

\vspace*{2cm}

{\bf Abstract}
\end{center}

We consider the high energy behaviour of the amplitudes for
production of leptons, quarks, Higgs bosons, sleptons, squarks, gauge
bosons, charginos and neutralinos at lepton colliders. 
We concentrate our discussion on the terms arising at one loop 
which grow logarithmically with the energy, typically 
$[a~ln{s\over m^2}-ln^2{s\over m^2}]$. We show that in each of the
above processes the coefficient "a" reflects in a remarkable 
way the basic gauge and Higgs structure of the underlying 
interactions. A comparison with experiments at future colliders
should thus provide a clean way to test the validity of the MSSM 
structure.

\setcounter{page}{1}

 \newpage

It is by now well-known that the electroweak radiative corrections
to standard processes increase strongly  with the center of 
mass energy $\sqrt{s}$.
This arises already at the one loop level
due to the presence 
of large double(DL) and single(SL) logarithms
${\alpha\over\pi} ln^2{s\over m^2}$  , 
${\alpha\over\pi}ln{s\over m^2}$,
\cite{Sud1,Sud2,log}. In the TeV range such terms reach the
several tens of percent and 
should be easily measurable
(and analyzable) at future lepton colliders \cite{LC} whose
experimental accuracy should be at the few permille level.
\\

The relevance of these large logarithmic effects 
at high energy colliders has been stressed recently
in the process $e^+e^-\to f\bar{f}$
for both the SM  \cite{log} and the Minimal Supersymmetric
Standard Model
(MSSM) cases \cite{susylog,tgbeta},
and in the process of production of scalar pairs in the MSSM
  \cite{scalar,HH}.
The
processes $e^+e^-\to\gamma\gamma,~\gamma Z,~ZZ$ \cite{gZ}
has been analyzed in the same spirit as well as
$\gamma\gamma\to f\bar{f}$ \cite{LR}
measurable at photon-photon colliders.
The logarithmic structure of the processes
$e^+e^-\to W^+W^-,~H^+H^-,~H^0_aH^0_b,~\chi^+\chi^-,~\chi^0\chi^0$
has also been analysed and we extract some results
from a publication in preparation \cite{chi}.\\

In order for these analyses of the MSSM cases
to be already meaningful in the one TeV range,
a necessary condition is that of a light SUSY scenario
in which all the SUSY masses relevant for the considered 
process are supposed to be not heavier than a few hundred GeV.
This condition can be rescaled for higher energies
provided that $M_{SUSY} << \sqrt{s}$. However,
the large size of the effects in the several TeV range 
would require a treatment
(even approximate) of the higher order effects in order to obtain a
good theoretical prediction (at the one percent level). Such attempts 
have already started \cite{resum} and it has been 
claimed that DL as well as universal and angular dependent SL can 
be exponentiated, see also \cite{scalar} for an application
to the MSSM case. 
This means that the one loop structure is the basic ingredient 
on which we can concentrate our discussion.\\

Comparing the SM and the MSSM logarithmic effects in these various
processes, we were impressed by a number of recurrent, 
impressively simple differences. The purpose of this short note 
is that of presenting in a systematic way these differences and 
of discussing their intuitive, deep physical origins.\\

\noindent
{\bf The logarithmic terms at the one loop level}\\

At the one loop level the logarithmic terms appearing in $e^+e^-$
processes can be separated into
three categories, Parameter Renomalization (PR) terms, Universal
terms
and Angular dependent terms :\\

(1) \underline{the PR terms}\\

These are the well-known terms generated by the
gauge boson or gaugino self-energy contributions. They are
process dependent, but can be computed in a straightforward
way from the Born amplitude i.e.,

\begin{equation}
{1\over4\pi^2}[g^4_1\beta_1{dA^{Born}\over dg^2_1}+
g^4_2\beta_2{dA^{Born}\over dg^2_2}][-ln{s\over\mu^2}]
\end{equation}

\noindent
where $\beta_{1}={-5\over9}N_{fam}-{1\over24},~(
-~{5\over6}N_{fam}-{1\over4})$ and 
$\beta_{2}=-~{1\over3}N_{fam}+{43\over24},~(
-~{1\over2}N_{fam}+{5\over4})$ are the 
SM, (MSSM) values of the usual RG
functions associated to the gauge
couplings $g_{1,2}$.\\

(2) \underline{Universal terms}\\

Also called "Sudakov" terms, 
these terms appear to be typically of the form
$[a~ln{s\over m^2}-ln^2{s\over m^2}]$. They 
factorize the Born amplitude in a process independent 
and angular independent fashion. They are
specific of the quantum numbers of each external particle. In the
high energy range they are the dominant terms.\\

The single logs correspond to collinear singularities
and the double logs to a coincidence of collinear and soft 
singularities in the one loop diagrams.\\

Although the results were first obtained by canonical computations
of self-energy, triangle and box diagrams,
in the covariant $\xi=1$ gauge, or in an axial gauge,
the simplest way to obtain these log terms 
is through the splitting function formalism \cite{split}.
With the splitting functions
$${1+x^2\over 1-x},~{1-x\over2},~{2x\over1-x},~ {1\over2},~
2[{x\over 1-x}+{1-x\over x}+x(1-x)],~
{x^2+(1-x)^2\over2},~ x(1-x),$$ for 
$f\to g f$, $f\to s f$, $s\to g s$, $s\to f\bar f$,
$g\to gg$, $g\to f\bar f$, $g\to s\bar s$ (where $f$,$s$,$g$ 
represent fermions, scalars and gauge bosons), respectively,
and the addition of the parameter renormalization terms,
one immediately recovers the results of the diagrammatic
computations.\\

(3)\underline{ Angular dependent terms}\\

They are just residual parts of the soft-collinear singularity 
arising from box diagrams involving gauge boson
exchanges which generate $ln^2|x|$ terms 
($x\equiv t\simeq-~{s\over2}(1-cos\theta)$ or 
$u\simeq-~{s\over2}(1+cos\theta)$, being the Mandelstam
parameters).
After having extracted
the universal angular independent part $ln^2s$,
one remains with an angular dependent terms of the type

$$2ln{s\over m^2}ln{|x|\over s}+ln^2{|x|\over s}$$ 

\noindent
whose contribution constitutes an additional process-dependent
single log. There are only few such terms, which are all of pure
SM origin and have been computed for all considered
processes.\\

Among these 3 types of terms the richest structure is 
in the Sudakov part (2), on which we now
concentrate. We write this contribution as

\begin{equation}
A^{1~loop}=[1~+~{\alpha\over\pi}~c]~A^{Born}
\end{equation}
\noindent
This factorization applies to each external line of the process,
$c$ being a coefficient that depends on the nature of the external
particle, on the type of interaction and on the energy.

When the external particle is one member of
two mixed states ($i=1,2$; see the examples below), 
the above equation has to be written in a matrix form:

\begin{equation}
A^{1~loop}_i = \sum_{j}~ [\delta_{ij}+
{\alpha\over\pi}c_{ij}]~A^{Born}_j
\end{equation}\\

We now list a number of typical examples.\\

\begin{center}
{\bf chiral lepton or quark}~~ ($f_{L,R}$) \\
\end{center}
\noindent
{\bf in the SM}

\vspace{-0.5cm}

\begin{eqnarray}
&&c(f_{L,R})={1\over8}~[3ln{s\over M^2_W}-ln^2{s\over M^2_W}]
~[~{I_f(I_f+1)\over s^2_W}
+{Y^2_f\over4c^2_W}~]_{L,R}\nonumber\\
&&
+[-ln{s\over M^2}]~[{1\over 32s^2_WM^2_W}]~\{~
[m^2_t+m^2_b][\delta_{f,t_L}+\delta_{f,b_L}]
+2m^2_t\delta_{f,t_R}+2m^2_b\delta_{f,b_R}~\}
\end{eqnarray}
\noindent
{\bf in the MSSM}

\vspace{-0.5cm}

\begin{eqnarray}
&&c(f_{L,R})={1\over8}~[2ln{s\over M^2_W}-ln^2{s\over M^2_W}]
~[~{I_f(I_f+1)\over s^2_W}
+{Y^2_f\over4c^2_W}~]_{L,R}\nonumber\\
&&
+~[-ln{s\over M^2}]~[{1\over 32s^2_WM^2_W}]
\{~[2m^2_t(1+\cot^2\beta)+2m^2_b(1+\tan^2\beta)]
[\delta_{f,t_L}+\delta_{f,b_L}]
\nonumber\\
&&
+[4m^2_t(1+\cot^2\beta)]\delta_{f,t_R}+
[4m^2_b(1+\tan^2\beta)]\delta_{f,b_R}~\}
\end{eqnarray}
One recognizes the Yukawa part appearing for heavy quarks,
where $\beta$ is the mixing angle between the vacuum 
expectation values of the 
up and down Higgs chiral superfield (in standard notation 
$\tan\beta = v_u/v_d$). The scale $M$ which appears in the single
logs is in principle the value of the highest mass
running inside the corresponding loop. In the SM case it should be
the top quark mass; in the MSSM case it will be a heavy squark
or a chargino mass. We shall come back to this point in the final
discussion.\\

\begin{center}
{\bf transverse $W^{\pm}_T,~\gamma,~Z_T$ 
in the SM and in the MSSM}
\end{center}

\begin{equation}
c(W)=~{1\over 4s^2_W}[-ln^2{s\over M^2_W}]
\end{equation} 

\begin{equation}
c_{\gamma\gamma}=~{1\over4}[-ln^2{s\over M^2_W}]~~~~~~~~~
c_{ZZ}=~{c^2_W\over4s^2_W}[-ln^2{s\over M^2_W}]
~~~~~~~~~ c_{\gamma Z}=~{c_W\over4s_W}[-ln^2{s\over M^2_W}]
\end{equation} 

\newpage

\begin{center}
{\bf neutral Higgs and charged or neutral
Goldstones  in the SM}
\end{center}
\vspace{-0.5cm}

\begin{eqnarray}
c(H_{SM})=c(G^0)=c(G^{\pm})&=&
~{(1+2c^2_W)\over 32s^2_Wc^2_W}~[4ln{s\over M^2_W}-ln^2{s\over M^2_W}]
\nonumber\\
&&
+{3\over 16s^2_WM^2_W}~[m^2_t+m^2_b]~[-ln{s\over M^2}]
\end{eqnarray}

\noindent
The charged and neutral Goldstone states are equivalent at high energy
to the longitudinal $W^{\pm}_L$ and $Z_L$ components.

\begin{center}
{\bf sleptons or squarks in the MSSM~~~($\tilde{f}_{L,R}$)}
\end{center}

Same expression as for leptons and quarks in the MSSM.\\

{\bf charged and neutral Higgs bosons and Goldstones 
in the MSSM }\\

A first $2\times2$ matrix describes the 
($H^{\pm},G^{\pm}\equiv W^{\pm}_L$) set,
as well as the ($A^{0},G^0\equiv Z_L$) set:

\begin{equation}
c_{11}=
~{(1+2c^2_W)\over 32s^2_Wc^2_W}~[2ln{s\over M^2_W}-ln^2{s\over M^2_W}]
+{3\over 16s^2_WM^2_W}~[m^2_tcot^2\beta+m^2_btan^2\beta]
~[-ln{s\over M^2}]
\end{equation}
\begin{equation}
c_{22}=
{(1+2c^2_W)\over 32s^2_Wc^2_W}~[2ln{s\over M^2_W}-ln^2{s\over M^2_W}]
+{3\over 16s^2_WM^2_W}~[m^2_t+m^2_b]~[-ln{s\over M^2}]
\end{equation}
\begin{equation}
c_{12}=
{3\over 16s^2_WM^2_W}~
[m^2_t\cot\beta-m^2_b\tan\beta]~[-ln{s\over M^2}]
\end{equation}

A second matrix describes the ($H^{0},h^{0}$) set

\begin{eqnarray}
c_{H^{0}H^{0}}&=&
{(1+2c^2_W)\over 32s^2_Wc^2_W}~[2ln{s\over M^2_W}-ln^2{s\over M^2_W}]
\nonumber\\
&&
+{3\over
16s^2_WM^2_W}~[m^2_t \sin^2\alpha(1+\cot^2\beta)
+m^2_b \cos^2\alpha(1+\tan^2\beta)]~[-ln{s\over M^2_W}]
\end{eqnarray}
\begin{eqnarray}
c_{h^{0}h^{0}}&=&
{(1+2c^2_W)\over 32s^2_Wc^2_W}
~[2ln{s\over M^2_W}-ln^2{s\over M^2}]\nonumber\\
&&
+{3\over 16s^2_WM^2_W}~[m^2_t \cos^2\alpha(1+\cot^2\beta)
+m^2_b \sin^2\alpha(1+\tan^2\beta)]~[-ln{s\over M^2}]
\end{eqnarray}
\begin{equation}
c_{H^{0}h^{0}}=
{3\cos\alpha \sin\alpha \over 16s^2_WM^2_W}~
[m^2_t(1+\cot^2\beta)
+m^2_b(1+\tan^2\beta)]~[-ln{s\over M^2}]
\end{equation}
where $\alpha$ is the mixing angle between the neutral 
CP even physical Higgs 
bosons; at tree level, $\alpha$ is a simple combination 
of $\tan\beta$ and the masses of the 
neutral (CP even and odd) physical Higgs bosons.\\

\begin{center}
{\bf charginos $\chi^{+}_i$ in the MSSM} 
\end{center}

\vspace{-1cm}

\begin{eqnarray}
&&c_{\chi^{+}_i\chi^{+}_j}=
{1\over 4s^2_W}~[-ln^2{s\over M^2_W}]~(Z^{+}_{1i}Z^{+}_{1j}P_L
+Z^{-}_{1i}Z^{-}_{1j}P_R)+\nonumber\\
&& 
{(1+2c^2_W)\over 32s^2_Wc^2_W}~[~2ln{s\over M^2_W}-ln^2{s\over M^2_W}~]
~(Z^{+}_{2i}Z^{+}_{2j}P_L
+Z^{-}_{2i}Z^{-}_{2j}P_R)+\nonumber\\
&&
[-ln{s\over M^2}]~({3\over 16s^2_W M^2_W})~[m^2_t(1+\cot^2\beta)
~Z^{+}_{2i}Z^{+}_{2j}P_L+m^2_b(1+\tan^2\beta)
~Z^{-}_{2i}Z^{-}_{2j}P_R]
\end{eqnarray}

The mixing matrix elements $Z^{\pm}_{1i}$ correspond \cite{Rosiek} 
to the charged gaugino components and $Z^{\pm}_{2i}$ to the
charged higgsino components.\\

\begin{center}
{\bf neutralinos $\chi^{0}_i$ in the MSSM}  
\end{center}
\vspace{-1cm}

\begin{eqnarray}
&&c_{\chi^{0}_i \chi^{0}_j}=
{1\over 4s^2_W}~[-ln^2{s\over M^2_W}]~(Z^{N}_{2i}Z^{N}_{2j})+
\nonumber\\
&& 
[{(1+2c^2_W)\over 32s^2_Wc^2_W}]
~[~2ln{s\over M^2_W}-ln^2{s\over M^2_W}~]
~(Z^{N}_{4i}Z^{N}_{4j}+Z^{N}_{3i}Z^{N}_{3j})+
\nonumber\\
&&
[-ln{s\over M^2}]~({3\over 16s^2_WM^2_W})[m^2_t(1+\cot^2\beta)~
Z^{N}_{4i}Z^{N}_{4j}
+m^2_b(1+\tan^2\beta)~Z^{N}_{3i}Z^{N}_{3j}]
\label{neut}\end{eqnarray}

The mixing matrix elements $Z^{N}_{2i}$ correspond \cite{Rosiek} to the
neutral gaugino $(\tilde{W_3})$ components (there is no
contribution from the Bino $\tilde{B}$), and $Z^{N}_{3i},~Z^{N}_{4i}$
to the neutral higgsino components.\\

{\bf Discussion of the SM terms}\\

The $[-ln^2{s\over M^2}]$ terms, in which $M=M_W$ or $M_Z$,
arise from the coincidence of soft and collinear
singularities. They only appear (because of helicity
conservation vertices) in SM gauge terms. They correspond to the
term $1/(1-x)$ in the splitting function with emission of a gauge boson.
The minus sign is fixed by unitarity (positivity of the transition
probability).\par
These SM gauge terms contain also
a single $[ln{s\over M^2}]$ part arising from the 
remaining part of the
gauge splitting functions. For a fermion line one obtains
the combination  $[~3ln{s\over M^2}-ln^2{s\over M^2}~]$ 
and for a scalar line $[~4ln{s\over M^2}-ln^2{s\over M^2}~]$.
The factor $3$ or $4$ can be traced back to the spin nature of
the gauge vertices $f\bar f g$, $s\bar s g$ 
(where $g$ is a gauge boson)
and more technically to the Lorentz transformation from the
c.m. frame to the collinear frame of the usual
$1+\cos^2\theta$ and $\sin^2\theta$ distributions 
of fermion or scalar pairs.\par

We have also obtained SL of Higgs origin due to Yukawa couplings to
heavy quarks. They appear both in heavy quark production processes
and in other final states involving Higgs and Goldstones
(where the heavy quarks 
contribute virtually). The minus
sign in front of the SL is also a consequence of unitarity.\\

{\bf Discussion of the SUSY terms and of the complete MSSM terms}\\

Additional single logarithmic terms $[-ln{s\over M^2}]$
arise from diagrams
involving scalar couplings of supersymmetric particles
(sfermions, charginos, neutralinos, charged or neutral Higgses).
They have also two different origins. First,  a gauge origin, 
with the same gauge couplings as in SM terms, so that the complete
MSSM combination is now $2ln-ln^2$.
Secondly, a Yukawa origin, but in this case the extended Higgs 
structure generates new contributions depending on the parameter 
$tan\beta$, and in the case of external $H^0,h^0$, also on the mixing
angle $\alpha$.
In the MSSM (and except for the $H^0,h^0$ case) $tan\beta$
is the only new SUSY parameter which enters the
asymptotic expressions, and leads to $m^2_t cot^2\beta$
and $m^2_b tan^2\beta$ terms.
The minus sign in front of the SL is
also a consequence of unitarity (positivity of the
corresponding transition probability). As already mentioned,
the scale $M$ which appears in these single logs is in principle 
the value of the highest mass
running inside the corresponding loop, but at logarithmic accuracy,
provided that one is especially interested in the slope in $log ~s$,
as suggested in Ref.\cite{tgbeta,scalar}, the choice of $M$
is harmless. \\

{\bf A list of benchmark features}\\

We now underline the benchmark features which arise from the
above results.\\

A first feature is that, \underline{for
lepton and quark production} the SM 
"fermion-gauge" combination
\begin{center}
 $[~3ln{s\over M^2_W}-ln^2{s\over M^2_W}~]$
becomes $[~2ln{s\over M^2_W}-ln^2{s\over M^2_W}~]$\\
\end{center} 
in the MSSM when gaugino terms are added. \\

The second feature is that, \underline{for
slepton and squark production} the SM "scalar-gauge" 
combination
\begin{center} $[~4ln{s\over M^2_W}-ln^2{s\over M^2_W}~]$ 
also becomes $[~2ln{s\over M^2_W}-ln^2{s\over M^2_W}~]$\\
\end{center} 
when the corresponding gaugino terms are added.\\
So the combination $[~2ln{s\over M^2_W}-ln^2{s\over M^2_W}~]$ 
appears to be the typical MSSM
combination of the whole
fermion-sfermion supermultiplet.\\

The transformation of $[~4ln{s\over M^2_W}-ln^2{s\over M^2_W}~]$ 
into $[~2ln{s\over M^2_W}-ln^2{s\over M^2_W}~]$ also occurs when
going from SM Higgs $H_{SM}$ or Goldstone 
$G^{\pm,0}\equiv W^{\pm},Z$ production, to MSSM 
charged or neutral Higgses $H^{\pm},~H^0,h^0,A^0$ or Goldstones.\\

For \underline{transverse gauge boson}
lines, only the quadratic term $[-ln^2{s\over M^2_W}]$ 
appears, both in the SM and in the MSSM.
This should provide a test of the assumed "minimal" gauge structure of
the MSSM, i.e. of
absence of additional (higher) gauge bosons.\\

In the case of chargino and neutralino production,
the $[~2ln{s\over M^2_W}-ln^2{s\over M^2_W}~]$
combination can also be found in the \underline{Higgsino components},
and the $[-ln^2{s\over M^2_W}]$ term  
in the \underline{gaugino
components}. This leads to 
an additional potential check of the assumed
supersymmetric nature of the
interactions of these particles which can be achieved by a
measurement of the production rate of the two charginos and of the four
neutralinos.\\

Finally, we consider the Yukawa terms which contribute
a  term $[-ln{s\over M^2}]$ in the production 
of \underline{heavy quarks}, \underline{heavy squarks},
\underline{Higgs bosons} and \underline{Goldstones}, 
\underline{charginos} and \underline{neutralinos}.
The interesting feature here is that the SM parameters 
$m^2_t$ and $m^2_b$
are, in the MSSM, replaced by terms that also contain 
the products $m^2_t \cot^2\beta$
and  $m^2_b \tan^2\beta$. For large values of $\tan^2\beta$
this would provide a genuine possibility of measuring 
this fundamental parameter,
as already stressed in \cite{tgbeta,scalar, tgbnote}.\\

The general conclusion that can be drawn from our 
analysis is that the genuine SUSY electroweak Sudakov 
logarithmic structure differs
from the corresponding one met in the SM in very simple and specific
ways. This suggest the following strategy.\par
Through the measurements of the coefficients
of the DL and SL, the production of usual particles 
(leptons, quarks, gauge bosons)
should provide global tests of the SM gauge and Higgs structure.
The spirit of these tests is similar to the one 
which motivates the
high precision tests with $g-2$ or $Z$ peak measurements.
If new particles, candidates for supersymmetry, exist, 
departures from SM predictions should appear and one 
should then compare with MSSM predictions for these log
coefficients. This can be done both for the production
of usual particles and for the production of the
new states (sleptons, light or heavy squarks, charged and neutral
Higgses, charginos, neutralinos). This should allow to check if
the MSSM description is satisfactory or if modifications or
extensions are needed (higher gauge bosons, more Higgses, ....,
or different forms of New Physics).\\

We would like to conclude by saying 
that even through the production of usual particles,
"virtual" supersymmetry would have a "reality" at future accelerators, 
since it exhibits peculiar "logarithmic fingerprints" that would 
be observable . Roughly, one might be tempted
to summarize these results via a rough "thumb rule", sounding like: 
\begin{center}
{\bf "0,1,2,3,4.....
count the logs and check supersymmetry"}.\\
\end{center}

\end{document}